\documentclass[10 pt,final,journal,letterpaper,oneside,onecolumn]{IEEEtran}
\usepackage{graphicx}
\usepackage{amssymb,amsmath,amsfonts}
\usepackage{suffix}
\usepackage{amssymb}
\usepackage{mathtools}
\usepackage{cite}  
\usepackage{amsthm}
\usepackage[utf8]{inputenc}
\usepackage[english]{babel}
\usepackage{epstopdf}
\usepackage{color}
\usepackage{times}
\usepackage{fancyhdr,graphicx,amsmath,amssymb}
\usepackage{algorithmicx}
\usepackage[ruled]{algorithm}
\usepackage{algpseudocode}
\ifCLASSINFOpdf
\else
\fi

\begin{document}
%
\title{Efficient Power-Splitting and Resource Allocation for Cellular V2X Communications}

\author{Furqan Jameel, Wali Ullah Khan, Neeraj Kumar, and Riku J\"antti \thanks{Furqan Jameel and Riku J\"antti are with the Department of Communications and Networking, Aalto University, 02150 Espoo, Finland. (email: furqanjameel01@gmail.com and riku.jantti@aalto.fi).

Wali Ullah Khan is with School of Information Science and Engineering, Shandong University, Qingdao 266237, China. (email: waliullahkhan30@gmail.com)

Neeraj Kumar is with the Department of Computer Science and Information Engineering, Asia University, Taiwan. (email: neeraj.kumar@thapar.edu).
}}%
\markboth{ACCEPTED IN IEEE TRANSACTIONS ON INTELLIGENT TRANSPORTATION SYSTEMS}%
{\MakeLowercase{\textit{et al.}}: ACCEPTED IN IEEE TRANSACTIONS ON INTELLIGENT TRANSPORTATION SYSTEMS}
\maketitle

\begin{abstract}
The research efforts on cellular vehicle-to-everything (V2X) communications are gaining momentum with each passing year. It is considered as a paradigm-altering approach to connect a large number of vehicles with minimal cost of deployment
and maintenance. This article aims to further push the state-of-the-art of cellular V2X communications by providing an optimization framework for wireless charging, power allocation, and resource block assignment. Specifically, we design a network model where roadside objects use wireless power from RF signals of electric vehicles for charging and information processing. Moreover, due to the resource-constraint nature of cellular V2X, the power allocation and resource block assignment are performed to efficiently use the resources. The proposed optimization framework shows an improvement in terms of the overall energy efficiency of the network when compared with the baseline technique. The performance gains of the proposed solution clearly demonstrate its feasibility and utility for cellular V2X communications.    
\end{abstract}


\IEEEpeerreviewmaketitle

\section{Introduction}
Recent breakthroughs in automated driving technologies and electric engines are expected to revolutionize the intelligent transportation systems (ITSs), in terms of on-road safety services, enriched travel experiences, and transportation efficiency \cite{storck20195g}. Although fully autonomous vehicles are still being developed, issues like collision warning/avoidance, inter-vehicle cooperation, and automated platooning need more improvements. In this regard, vehicle to everything (V2X) communications is expected to solve most of these issues through interaction, coordination, and cooperation with the devices on the road \cite{jameel2019performance,kumar2015bayesian}. Some works for onboard sensing and computer vision share the same objective but show little performance gains. Specifically, these ambitious projects are constrained by many technical challenges like training data sets, machine defects, poor processing capabilities and limiting sensing range. V2X communication is expected to break these barriers through cooperation among vehicles via wireless communications \cite{ren2015power,khan2020multiobjective,kumar2015intelligent}. It also aims to empower cross-discipline studies on computer vision and onboard sensing through the integration of real-time global information \cite{jameel2020efficient}. 

\subsection{Related Work}
\begin{figure*}[!htp]
\centering
\begin{tabular}{c}
\includegraphics[scale=.45]{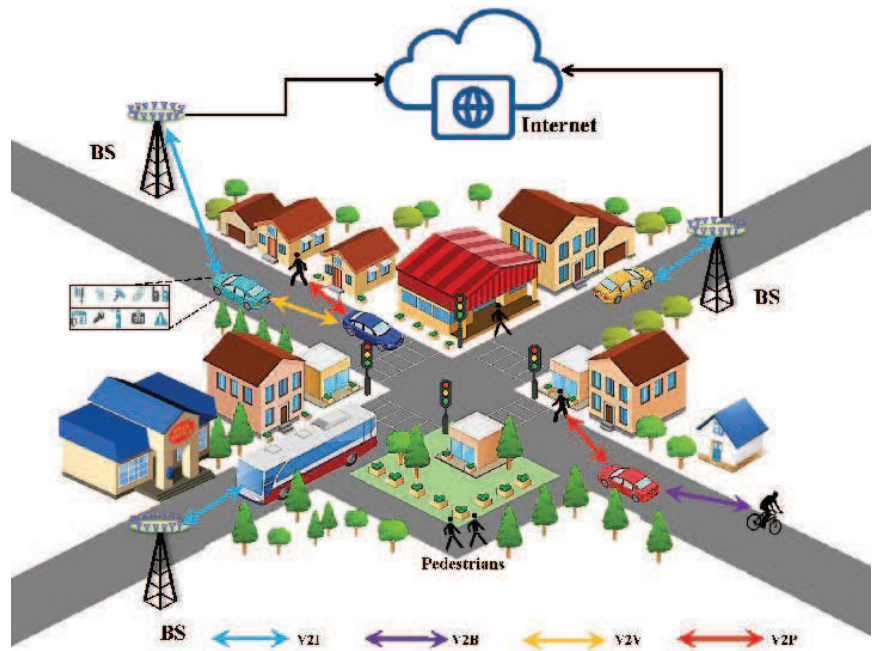}
\end{tabular}
\caption{An illustration of cellular V2X communications. The vehicles are connected to the cellular BS for transferring the data over the Internet. On the road, the vehicles are connected to the roadside objects such as sign-posts, bicycles etc.}
\label{fig1}
\end{figure*}

The resource allocation problem in V2X communications has recently been studied by researchers in academia and industry. This was not the case as earlier studies focused mostly on the modeling and performance evaluation of vehicular networks \cite{jameel2018performance,raza2020spatial,jameel2017performance}. For instance, the authors of \cite{mei2018latency} proposed a resource allocation solution to improve the service rate of the network users. Zhou \emph{et al.} in \cite{zhou2017social} used the game theory approach to improve network performance. More specifically, a two-stage auction matching resource allocation technique was presented by the authors. One of the critical requirements of roadside objects is sensing the environment in real-time and forward the collected data to the nearby vehicles. In this way, the vehicles could perform analysis for decision making. \textcolor{black}{Similarly, the vehicle may send its whereabouts, along with other useful information to the roadside objects.} Thus, the roadside objects should be equipped with proper hardware and need to be active all the time to support such types of downlink information. For the case of uplink communications, the authors of \cite{ren2015power} proposed a resource allocation scheme for a freeway V2V communication for maximizing the sum-rate of the wireless links. However, in this work, the authors neglected the transmission requirements of roadside objects. Moreover, these techniques are limited to the optimization of a single time slot and neglect the optimization of the parameters over long-time constraints. \textcolor{black}{Wang \emph{et al.} in \cite{wang2018cellular} focused on the energy-sensing aspect of C-V2X networks and aimed at improving the spectrum sharing abilities of such networks. They did so by constructing the interference model and by allocating resources using matching theory. Similar work was done by the authors of \cite{abbas2018novel} for latency minimization in C-V2X. They proposed a greedy link selection algorithm that minimizes the latency of vehicles using the 802.11p protocol. The results show the effectiveness of their approach as opposed to conventional techniques.}

The authors of \cite{liang2017resource} aimed to maximize the reliability and ergodic capacity of the cellular V2X communication network for a multi-lane freeway scenario. The main resource allocation constraint was outage probability, whereby, orthogonal resource blocks were allocated to all the users. They applied the Hungarian method and showed that their scheme improves the performance of the network. The authors of \cite{zhang2016resource} considered a broadcast scenario and maximized the number of concurrent V2V transmissions. The main constraints were on the signal power while the non-orthogonal resource blocks were allocated for the communication between vehicles and roadside objects. They applied the Perron Frobenius theory and showed that their resource allocation framework improves the performance of the network. MmWave communication has already shown a considerable advantage over sub-6 GHz networks \cite{zhang2019optimal,jameel2018wireless}. They evaluated the performance of different state-of-the-art techniques of mmWave vehicular communications and outlined their respective advantages and disadvantages. Feng \emph{et al.} in \cite{feng2020joint} provided a mobile edge computing assisted framework for C-V2X communications. Their main focus was on enabling ultra-reliable low-latency communication in such networks via computation offloading and resource allocation. However, the vehicles on road were distributed randomly in their study which may not be practical in most settings. In another recent work \cite{xu2020cellular}, Xu \emph{et al.} proposed to use a weighted mode selection approach to improve the coverage probability of the C-V2X communications. They compared their results to three different benchmark scheme. It was shown that their proposed solution outperforms the benchmark schemes in a comprehensive manner. The authors of \cite{wei2017resource,liang2017spectrum} proposed resource allocation techniques for maximizing the throughput of the V2V systems over the orthogonal resource block allocation among users. They, respectively, used hypergraph matching theory and Hungarian method to solve the optimization problems in an urban and freeway communication scenario. More recently, the authors of \cite{masmoudi2019efficient} considered a unicast communication paradigm for guaranteeing the sum-rate and reliability constraints. The authors used signal power and buffer size as the main constraints for V2X communication in a freeway scenario. Similarly, the authors of \cite{masmoudi2018mixed} improved reliability and reduce the latency of V2X communication for safety applications. More specifically, they optimized the network performance against signal power and delay constraints. They considered orthogonal resource block allocation among users communicating to the vehicle on a freeway scenario. Their proposed mixed traffic sharing and resource allocation algorithm shows noteworthy performance gains due to the efficient utilization of resources.

\subsection{Motivation and Contribution}

Though recent developments in cellular V2X communication have resulted in noteworthy improvements, the energy efficiency aspect of such networks has received little attention. As discussed previously, most of the existing studies focus on short-term optimization gains while neglecting the system performance improvements in the long-term. The resource allocation performed in a greedy way results in long-term performance degradation. Moreover, with the involvement of ultra-low powered roadside objects and electric vehicles, it is becoming ever more necessary to improve the energy efficiency of the overall V2X network without compromising the quality of service requirements. Thus, motivated by this prospect, our article makes the following contributions to the state-of-the-art:
\begin{enumerate}
\item A dynamic wireless-power transmission system model has been developed, whereby, electric vehicles communicate to the wireless-powered roadside objects. \textcolor{black}{The roadside objects can be wirelessly charged by the RF signal received from the desired electric vehicles and the interfering signal while making use of a power-splitting (PS) receiver design for information processing and wireless charging.}
\item To make the best tradeoff between the transmission power-saving and total capacity enhancement, an energy efficiency problem is formulated for the entire V2X network. \textcolor{black}{Different performance constraints of roadside objects are also taken into account.}
\item An optimization framework for PS adjustment, electric vehicle power allocation, and resource block assignment is provided. The optimization framework is compared with the baseline technique which shows the superiority of the proposed method.
\end{enumerate}

\subsection{Organization}

The rest of the paper is organized as follows. Section II provides a detailed description of the system model and problem formulation. In Section III, step-by-step details of the proposed optimization framework are provided. Section IV gives simulation results and relevant discussion. Finally, Section V provides some concluding remarks and future work. The list of the notations used in this paper is given in Table \ref{tab1}.

\section{System Model and Problem Formulation}

In this section, we provide details of the system model along with the discussion on problem formulation.

\subsection{System Model}
\begin{figure}[!htp]
\centering
\begin{tabular}{c}
\includegraphics[scale=.4]{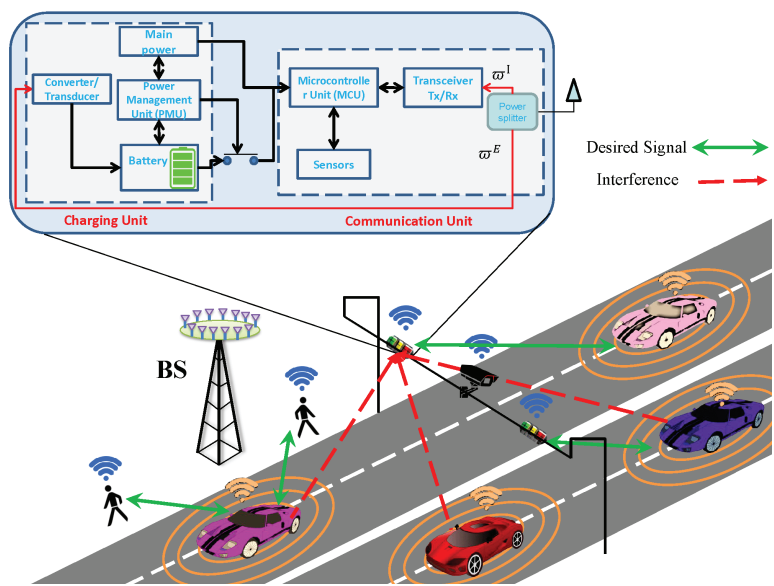}
\end{tabular}
\caption{Two-lane system model with multiple electric vehicles and roadside objects. The roadside objects are equipped with wireless power and information processing receivers.}
\label{fig2}
\end{figure}

We consider a two-lane road having $N$ electric vehicles with limited energy storage capacity. The vehicles are considered to be communicating with the roadside objects (e.g. pedestrians with wearables or road signs) with the bandwidth $W$ and share the same resource block $B$ over the time intervals $L$. \textcolor{black}{Broadly speaking, the BS mainly assists in the allocation of resources and ensures the reliability of services. It is also responsible for the allocation of subchannels for efficient V2X communications during each transmission time interval \cite{nardini2018cellular}.} Moreover, at any time interval $l$, the $i$-th electric vehicle communicates with the $J_{i}(l)$ roadside objects. Both the electric vehicle and the roadside objects are assumed to be equipped with single omnidirectional antennas. The communication links among electric vehicles and that between vehicles and roadside objects are assumed to follow Rayleigh distributions. 

\begin{table}[!htp]
\centering
\caption{List of commonly used notations.}
\label{tab1}
\begin{tabular}{|l|l|}
\hline
\textbf{Notation} & \textbf{Definition}                    \\ \hline \hline
$g$               & Channel gain                           \\ \hline
$\varpi^I$        & Information processing ratio           \\ \hline
$\varpi^E$        & Wireless charging ratio                \\ \hline
$\eta$            & Power conversion efficiency            \\ \hline
$N$               & Total electric vehicles                \\ \hline
$B$               & Total resource blocks                  \\ \hline
$P$               & Transmit power of electric vehicle     \\ \hline
$\beta,\delta,\tau,\theta,\lambda,\pi$                & Lagrangian multipliers                 \\ \hline
$N_0$             & Additive white Gaussian noise variance \\ \hline
$\sigma$          & Roadside object association indicator  \\ \hline
$L$               & Time intervals                         \\ \hline
$W$               & Bandwidth                              \\ \hline
$C$               & Charging capacity of roadside object   \\ \hline
N\_p              & Signal processing noise varience       \\ \hline
$E^V$             & Total power consumption at vehicle     \\ \hline
$E^{RO}$          & Power consumed at roadside object      \\ \hline
$\Lambda$         & Utility function                     \\ \hline
$R$               & Total achievable capacity              \\ \hline
$E$               & Total power consumption                \\ \hline
\end{tabular}
\end{table}

All the roadside objects are assumed to be equipped with wireless charging hardware. Specifically, the hardware design of roadside objects make use of PS mechanism such that the received RF power at the antenna is split into two streams for information processing and wireless charging. Let us denote the PS ratios for wireless charging and information processing as $\varpi _{i,j,r}^{E}(l)$ and $\varpi _{i,j,r}^{I}(l)$, when $i$-th electric vehicle communicates to $j$-th roadside object over the $r$-th resource block. Hence, the wireless charging and information processing policies can be, respectively, given as 
\begin{align}
\varpi ^{E}=\lbrace \varpi _{i,j,r}^{E}(l)\in \lbrack 0,1\rbrack , \forall 
i,\forall j,\forall r\rbrace,
\end{align}
and 
\begin{align}
\varpi ^{I}=\lbrace \varpi _{i,j,r}^{I}(l)\in \lbrack 0,1\rbrack , \forall i,\forall j,\forall 
r\rbrace .
\end{align}

\subsubsection{Wireless Power Transmission}

For the communication with roadside objects, the electric vehicle consumes the power of the battery. Thus, the power consumption of the $i$-th electric vehicle in time interval $l$ can be written as
\begin{align}
E_{i}^{V}(l)= P^{V}+\sum_{j=1}^{J_{i}(l)}{\sum_{r=1}^{B}{{\sigma}_{i,j,r}(l)}P_{i,j,r}^{}(l)},
\end{align}
where $P^{V}$ is the static power consumed by the vehicle and $P_{i,j,r}^{}(l)$ is the transmitted power from $i$-th electric 
vehicle to $j$-th roadside object on $r$-th resource block. Furthermore, ${\sigma}_{i,j,r}(l) \in \lbrace 0,1\rbrace $ denotes the binary indicator which indicates that the electric vehicle is communicating to roadside object using $r$-th resource block. 

\textcolor{black}{Since the roadside objects can charge their batteries using the received power from the RF signal, it is possible to improve their lifecycle.} Moreover, it is also worth mentioning that the roadside objects not only charge their battery from the desired electric vehicle, but also from the interfering signals from the other electric vehicles. \textcolor{black}{When the $j$-th roadside object communicates to the $i$-th electric vehicle communicates, the amount of wireless charging at time interval $l$ can be given as}
\begin{align}
C_{i,j}(l)&=\sum_{r=1}^{B}{\eta \varpi _{i,j,r}^{E}(l){\sigma}_{i,j,r}(l)\lbrace 
I_{i,j,r}+N_{0}W\rbrace }\nonumber \\
&+\sum_{r=1}^{B}{\eta \varpi_{i,j,r}^{E}(l){\sigma}_{i,j,r}(l)P_{i,j,r}^{}g_{i,j,r}(l)},
\end{align}
where $0<\eta <1$ is the energy conversion efficiency, $N_{0}$ is 
the thermal noise of the roadside object, and $g_{i,j,r}(l)$ is the 
wireless channel between the $i$-th electric vehicle and $j$-th 
roadside object. Also, $I_{i,j,r}$ represents the interference from 
the other electric vehicles to the $j$-th roadside object, which can 
be given as
\begin{align}
I_{i,j,r}=\sum_{v\ne 
i}^{N}{\sum_{j^{*}=1}^{J_{i^{*}}(l)}{{\sigma}_{i^{*},j^{*},r}(l)}}P_{i^{*},j^{*},r}^{}g_{i^{*},j^{*},r}(l).
\label{eq_05}
\end{align}

The power consumption of the $j$-th roadside object at any time interval $l$ can be, thus, given as
\begin{align}
E_{i,j}^{RO}(l)=\max \lbrace P^{RO}-C_{i,j}(l),0\rbrace,
\label{eq_6}
\end{align}
where $P^{RO}$ denotes the consumption of power during data reception at the roadside object. 

Now, the total power consumption of the V2X network can be written as
\begin{align}
E^{}= \sum_{l=1}^{L}{\sum_{i=1}^{N}{\sum_{j=1}^{J_{i}(l)}{E_{i,j}^{RO}(l)}}}+\sum_{l=1}^{L}{\sum_{i=1}^{N}{E_{i}^{V}(l)}}.
\end{align}

\subsubsection{Information Processing}

During each time interval $l$, \textcolor{black}{the $i$-th electric vehicle} communicates to the $j$-th roadside object using $r$-th resource block. \textcolor{black}{It is worth pointing out that in the considered setup, each vehicle communicates to the roadside object independently. This means that there is no transfer of multicast messages and the BS allocate resources to roadside objects. Yet, due to simultaneous transmission of other electric vehicles in the range, the desired roadside object may receive interference from other vehicles.} Therefore, the signal-to-interference-and-noise ratio (SINR) can be given as 
\begin{align}
\gamma _{i,j,r}(l)=\frac{\varpi 
_{i,j,r}^{l}(l)P_{i,j,r}^{}g_{i,j,r}(l)}{N_{p}+\varpi 
_{i,j,r}^{l}(l)(I_{i,j,r}+N_{0}W)}
\end{align}
where $N_{p}$ represents the signal processing noise \cite{zhang2013mimo} at the roadside object and $I_{i,j,r}$ is given in (\ref{eq_05}).

Now, over the time intervals $L$, the total achievable capacity of the V2X network can be expressed as
\begin{align}
R=\sum_{l=1}^{L}{\sum_{i=1}^{N}{\sum_{j=1}^{J_{i}(l)}{\sum_{r=1}^{B}{W{\sigma}_{i,j,r}(l)}\log 
_{2}(1+\gamma _{i,j,r}(l))}}}.
\end{align}

\subsection{Problem Formulation} 

In this article, our main objective is to improve the overall energy efficiency of the V2X network, given the performance constraints of electric vehicle and roadside objects. The energy efficiency is defined as the ratio between total capacity of the V2X network and the total consumed power of the network. In this regard, we aim to jointly adjust the PS ratios $(\varpi ^{E},\varpi ^{I})$, resource block assignment $\sigma=\lbrace {\sigma}_{i,j,r}(l)| \forall i, \forall j,\forall r\rbrace $, and power allocation $P=\lbrace P_{i,j,r}^{}(l)| \forall i, \forall j,\forall r\rbrace $. Thus, the optimization problem can be written as
\begin{align}
\hspace*{-3cm} \max _{\varpi ,{\sigma},P}\Lambda(\varpi ,{\sigma},P)
\label{eq_10}
\end{align}
\begin{align}
\text{s.t.} &\textbf{C1}: \sum_{r=1}^{B}{W{\sigma}_{i,j,r}(l)}\log _{2}(1+\gamma _{i,j,r}(l))\ge \Pi _{i,j}^{\min }(l) \forall i, j,l \nonumber \\
&\textbf{C2}: \varpi _{i,j,r}^{I}(l)+\varpi _{i,j,r}^{E}(l)=1, \forall i,j, l, r \nonumber \\
&\textbf{C3}: \sum_{j=1}^{J_{i}(l)}{{\sigma}_{i,j,r}(l)\le 1}, \forall i,l, r \nonumber \\
&\textbf{C4}: C_{i,j}(l)\ge P_{i,j}^{\min }(l), \forall i, j,l \nonumber \\
&\textbf{C5}: \sum_{j=1}^{J_{i}(l)}{\sum_{r=1}^{B}{{\sigma}_{i,j,r}(l)P_{i,j,r}^{}(l)\le P_{i}^{T}}}, \forall i,l \nonumber \\
&\textbf{C6}: 0<\varpi _{i,j,r}^{I}(l),\varpi _{i,j,r}^{E}(l)<1, \forall i,l, j \nonumber
\end{align}
where $\Lambda(\varpi ,{\sigma},P)=\frac{R(\varpi ,{\sigma},P)}{E^{}(\varpi ,{\sigma},P)}$, \textbf{C1} is the quality of service constraint on the roadside object such that $\Pi _{i,j}^{\min }(l)$ denotes the minimum required data rate at any time interval $l$. The \textbf{C2} ensures that the sum of PS ratios for information processing and wireless charging remains 1. The \textbf{C3} ensures that the resource block $r$ is assigned to only one roadside object at any time interval $l$ to avoid any interference among the roadside objects of $i$-th electric vehicle. In \textbf{C4}, $P_{i,j}^{\min }(l)$ represents the minimum charging capacity required for operation of the roadside object. The \textbf{C5} puts constraint on the maximum transmit power such that at any time interval $l$, the transmit power does not exceed $P_{i}^{T}$. The \textbf{C6} is for ensuring the boundary limits on the PS ratios.

\section{Proposed Energy Efficient Solution}

From (\ref{eq_10}), one can observe that the problem is non-convex due to the existence of the fraction form of the objective function and the binary indicators. It is very difficult to find a solution to such a mixed-integer non-convex problem. Therefore, we transform the problem to address the non-convexity and non-linearity of the original problem. 

Let us first relax the binary indicator by considering ${\sigma}_{i,j,r}(l)$ 
as a continuous variable. Thus, it can be expressed as
\begin{align}
0\le {\sigma}_{i,j,r}(l)\le 1.
\end{align}

Now, (\ref{eq_6}) can be rewritten as
\begin{align}
E_{i,j}^{RO}(l)=P^{RO}-C_{i,j}(l),
\end{align}
such that $C_{i,j}(l)\le P^{RO}, \forall i,j,l$.

We first address the non-convexity of the optimization problem due to interference expression. Thus, following the approach of \cite{papandriopoulos2006low}, we use the approximation of transmission rate expressed as

\begin{align}
\underbrace{\sum_{r=1}^{B}{W{\sigma}_{i,j,r}(l)}\log _{2}(1+\gamma 
_{i,j,r}(l))}_{\Pi_{i,j} (l)}&\le \sum_{r=1}^{B}W{\sigma}_{i,j,r} \lbrace a_{i,j,r}(l) \nonumber \\
&\times \log _{2}\gamma _{i,j,r}(l)+b_{i,j,r}(l)\rbrace ,
\end{align}
where the values of the coefficients $(a_{i,j,r}(l),b_{i,j,r}(l))$ can 
be obtained iteratively. For the iteration index $c\ge 1$, the values 
of $a_{i,j,r}(l)$ and $b_{i,j,r}(l)$ can be, respectively, obtained 
as

\begin{align}
a_{i,j,r}^{(c)}=\frac{\gamma _{i,j,r}^{(c-1)}}{\gamma 
_{i,j,r}^{(c-1)}+1},
\end{align}
and 
\begin{align}
b_{i,j,r}^{(c)}=\log (1+\gamma _{i,j,r}^{(c-1)})-\frac{\gamma 
_{i,j,r}^{(c-1)}}{\gamma _{i,j,r}^{(c-1)}+1}\log \gamma 
_{i,j,r}^{(c-1)}.
\end{align}

\textcolor{black}{Besides this, to transform the problem into a concave optimization problem, we consider using a constant value of interference. Using a dynamic value may make the problem more complex and infeasible to solve comprehensively. Thus, we select a worst case scenario for the level of interference in the network.} Now the transmission rate can be approximated as
\begin{align}
\tilde{\Pi }_{i,j}(l)=\sum_{r=1}^{B}{W{\sigma}_{i,j,r}\lbrace a_{i,j,r}(l).\log 
_{2}\tilde{\gamma }_{i,j,r}(l)+b_{i,j,r}(l)\rbrace }
\end{align}
where $\tilde{\gamma }_{i,j,r}(l)=\frac{\varpi 
_{i,j,r}^{l}(l)P_{i,j,r}^{}g_{i,j,r}(l)}{N_{p}+\varpi 
_{i,j,r}^{l}(l)(\tilde{I}_{i,j,r}+N_{0}W)}$. Furthermore, the value of $
\tilde{I}_{i,j,r}$ is kept at a tolerable interference level.

\textcolor{black}{Note that hardware impairments are inherent part of most wireless devices. Due to this reason, it is very difficult to estimate the impact of such impairments on performance of a large-scale V2X network. Thus, to simplify the problem, we consider an ideal receiver at the roadside object. This receiver is assumed to fully harvest the energy received from the desired electric vehicle \cite{ng2013wireless}.} The total amount of wireless charging is now upper bounded as
\begin{align}
\tilde{C}_{i,j}(l)&=\sum_{r=1}^{B}{\eta \varpi 
_{i,j,r}^{E}(l){\sigma}_{i,j,r}(l)\lbrace \tilde{I}_{i,j,r}+N_{0}W\rbrace } \nonumber \\
&+\sum_{r=1}^{B}{\eta {\sigma}_{i,j,r}(l)P_{i,j,r}^{}g_{i,j,r}(l)},
\end{align}

Now we focus our attention on the non-linearity of the objective function. To provide a mathematically tractable solution, it is important to deal with the fraction form of the objective function. With this intent, we employ Dinkelbach method \cite{dinkelbach1967nonlinear} to transform the objective function. Let us first define the maximum energy efficiency of the V2X network as

\begin{align}
\Lambda^{*}=\frac{R(\varpi ^{*},{\sigma}^{*},P^{*})}{E^{}(\varpi ^{*},{\sigma}^{*},P^{*})}
\end{align}

where $\varpi ^{*},{\sigma}^{*},P^{*}$ are optimal PS ratios, 
resource block assignment, and power allocation. In principle, the 
maximum energy efficiency $\Lambda^{*}$ can be achieved when 
\begin{align}
&\max _{\varpi ,{\sigma},P}R(\varpi ,{\sigma},P)-\Lambda^{*}E^{}(\varpi ,{\sigma},P)\nonumber \\
&=R(\varpi ^{*},{\sigma}^{*},P^{*})-\Lambda^{*}E^{}(\varpi ^{*},{\sigma}^{*},P^{*})=0
\end{align}

Thus, the original problem can now be written as 
\begin{align}
\max _{\varpi ,{\sigma},P}R(\varpi ,{\sigma},P)-{\Lambda}E^{}(\varpi ,{\sigma},P).
\label{eq_20}
\end{align}

For a given parameter $\Lambda$, the expression in (\ref{eq_20}) can be solved iteratively. Now, the objective function in (\ref{eq_10}) can be transformed as
\begin{align}
\hspace*{-2cm} \max _{\varpi ,{\sigma},P}\tilde{R}(\varpi ,{\sigma},P)-\Lambda^{*}\tilde{E}^{}(\varpi ,{\sigma},P) 
\label{eq_21}
\end{align}
\begin{align}
\text{s.t.} & \textbf{C1}: \tilde{\Pi }_{i,j}(l)\ge \Pi _{i,j}^{\min }(l) \forall i, j,l \nonumber \\
& \textbf{C2}: \varpi _{i,j,r}^{I}(l)+\varpi _{i,j,r}^{E}(l)=1, \forall i,j, l, r \nonumber \\
& \textbf{C3}: \sum_{j=1}^{J_{i}(l)}{{\sigma}_{i,j,r}(l)\le 1}, \forall i,l, r \nonumber \\
& \textbf{C4}: \tilde{C}_{i,j}(l)\ge P_{i,j}^{\min }(l), \forall i, j,l \nonumber \\
& \textbf{C5}: \sum_{j=1}^{J_{i}(l)}{\sum_{r=1}^{B}{{\sigma}_{i,j,r}(l)P_{i,j,r}^{}(l)\le P_{i}^{T}}}, \forall i,l \nonumber \\
& \textbf{C6}: 0<\varpi _{i,j,r}^{I}(l),\varpi _{i,j,r}^{E}(l)<1, \forall i,l, j \nonumber \\
& \textbf{C7}: \tilde{C}_{i,j}(l)\le P^{RO}, \forall i,j,l \nonumber 
\end{align}
where $\tilde{R}(\varpi,{\sigma},P)=\sum_{l=1}^{L}{\sum_{i=1}^{N}{\sum_{j=1}^{J_{i}(l)}{{\Pi 
}_{i,j}(l)}}}$ and $\tilde{E}^{}(\varpi ,{\sigma},P)=\sum_{l=1}^{L}{\sum_{i=1}^{N}{\sum_{j=1}^{J_{i}(l)}{\tilde{E}_{i,j}^{RO}(l)}}}+\sum_{l=1}^{L}{\sum_{i=1}^{N}{E_{i}^{V}(l)}}
$. Here, due to transformation, $\tilde{E}_{i,j}^{RO}(l)$ is expressed as
\begin{align}
\tilde{E}_{i,j}^{RO}(l)=P^{RO}-\tilde{C}_{i,j}(l).
\end{align}

Note that the optimization problem now is standard concave maximization problem and can be solve using Lagrangian dual method. Let us first define the policies of PS ratios and power allocation on each resource block. \textcolor{black}{Hence, the new power allocation term can be expressed as
\begin{align}
\hat{P}_{i,j,r}^{ }(l)=P_{i,j,r}^{ }(l){\sigma}_{i,j,r}(l).
\end{align}}

\textcolor{black}{In a similar manner, the new PS factors can be written as
\begin{align}
\hat{\varpi }_{i,j,r}^{I}(l)=\varpi _{i,j,r}^{I}(l){\sigma}_{i,j,r}(l) \\
\hat{\varpi }_{i,j,r}^{E}(l)=\varpi _{i,j,r}^{E}(l){\sigma}_{i,j,r}(l) 
\end{align}}

Now, the Lagrangian function to solve (\ref{eq_21}) is expressed as
\begin{align}
L(\Omega ,\varpi ,{\sigma},P)&=\tilde{R}(\varpi ,{\sigma},P)-\Lambda^{*}\tilde{E}^{ }(\varpi ,{\sigma},P) \nonumber \\
&+\sum_{l=1}^{L}{\sum_{i=1}^{N}{\sum_{j=1}^{J_{i}(l)}{\beta _{i,j}(l)\biggl \lbrace \tilde{\Pi }_{i,j}(l)-\Pi _{i,j}^{\min }(l)\biggr \rbrace }}}\nonumber \\
&+\sum_{l=1}^{L}\sum_{i=1}^{N}\sum_{j=1}^{J_{i}(l)}\sum_{r=1}^{B}\delta _{i,j,r}(l)\lbrace {\sigma}_{i,j,r}(l) \nonumber \\
&-\hat{\varpi }_{i,j,r}^{I}(l)-\hat{\varpi }_{i,j,r}^{E}(l)\rbrace  \nonumber \\
&+\sum_{l=1}^{L}{\sum_{i=1}^{N}{\sum_{r=1}^{B}{\tau _{i,r}(l)}}\biggl \lbrace 1-\sum_{j=1}^{J_{i}(l)}{{\sigma}_{i,j,r}(l)}\biggr \rbrace } \nonumber \\
&+\sum_{l=1}^{L}{\sum_{i=1}^{N}{\sum_{j=1}^{J_{i}(l)}{\theta _{i,j}(l)}\biggl \lbrace \frac{\tilde{C}_{i,j}(l)}{\eta \hat{\varpi }_{i,j,r}^{E}}-\frac{P_{i,j}^{\min }(l)}{\eta \hat{\varpi }_{i,j,r}^{E}(l)}\biggr \rbrace }} \nonumber \\
&+\sum_{l=1}^{L}{\sum_{i=1}^{N}{\lambda _{i}(l)}\biggl \lbrace P_{i}^{T}-\sum_{j=1}^{J_{i}(l)}{\sum_{r=1}^{B}{\tilde{P}_{i,j,r}^{ }(l)}}\biggr \rbrace 
} \nonumber \\
&+\sum_{l=1}^{L}{\sum_{i=1}^{N}{\sum_{j=1}^{J_{i}(l)}{\pi _{i,j}(l)} \biggl \lbrace \frac{P^{RO}}{\eta \hat{\varpi }_{i,j,r}^{E}}-\frac{\tilde{C}_{i,j}(l)}{\eta \hat{\varpi }_{i,j,r}^{E}(l)}\biggr \rbrace }},
\end{align}
where the set of Lagrangian multipliers is given as $\Omega =\lbrace \beta _{i,j}(l),\delta _{i,j,r}(l),\tau _{i,r}(l),\theta _{i,j}(l),\lambda _{i}(l),\pi _{i,j}(l)\rbrace $. \textcolor{black}{Now, the Karush-Khun-Tucker (KKT) conditions can be applied to the problem in following manner \cite{khan2020efficient,khan2019joint}
\begin{align}
\frac{\partial{L(\Omega ,\mathbf{\varpi} ,{\sigma},P)}}{\partial \mathbf{\varpi}} \mid_{\mathbf{\varpi}=\mathbf{\varpi}^{*}} =0
\end{align}
where $\partial(.)$ indicate the partial derivative. After some straightforward mathematical steps, we can find the PS ratio as}
\begin{align}
\varpi _{i,j,r}^{I*}(l)=\biggl \lbrack \frac{\sqrt[]{\Xi _{i,j,r}(l)}-\ln 
(2)\delta _{i,j,r}(l)N_{p}}{2(\tilde{I}_{i,j,r}+N_{0}W)\ln (2)\delta 
_{i,j,r}(l)}\biggr \rbrack _{0}^{1}
\end{align}
where $ \lbrack h\rbrack _{0}^{1}=\biggl \lbrace \begin{matrix}
0 & h<0 & \\
h & 0\le h\le 1 & \\
1 & h>1 & \\
\end{matrix}
$ 

and 
\begin{align}
\Xi _{i,j,r}(l)=(\tilde{I}_{i,j,r}+N_{0}W)(1+\beta _{i,j}(l))4\ln 
(2)W \nonumber \\
\times N_{p}\delta _{i,j,r}(l)a_{i,j,r}(l)+(\ln (2)N_{p}\delta 
_{i,j,r}(l))^{2}
\end{align}

Given the value of $\varpi _{i,j,r}^{I*}(l)$ is known for information 
processing, the value of PS ratio for wireless charging can 
be written as
\begin{align}
\varpi _{i,j,r}^{E*}=1-\varpi _{i,j,r}^{I*}(l)
\end{align}

\textcolor{black}{In a similar manner, the policy for optimal power allocation can be obtained by taking partial derivative as
\begin{align}
\frac{\partial{L(\Omega ,\mathbf{\varpi} ,{\sigma},P)}}{\partial {P}} \mid_{{P}={P}^{*}} =0
\end{align}
Subsequently, after some simple mathematical manipulations, we obtain} 
\begin{align}
P_{i,j,r}^{ *}(l)=\max \biggl\lbrace \frac{a_{i,j,r}(l)W(1+\beta _{i,j}(l))}{\ln (2)\Theta _{i,j,r}(l)},0 \biggr \rbrace 
\end{align}
where 

\begin{align}
\Theta _{i,j,r}(l)=\Lambda +(\pi _{i,j}(l)-\theta _{i,j}(l)-\Lambda\eta )g_{i,j,r}(l)+\lambda _{i}(l).
\end{align}

For the case of resource block allocation, we use the concept of marginal benefits \cite{yu2002fdma}. Accordingly, for any $i$-th electric vehicle communicating to $j$-th roadside object, the resource block allocation policy can be given as

\begin{align}
{\sigma}_{i,j^{*},r}^{*}(l)=\biggl \lbrace \begin{matrix}
1, & j=\arg\max _{j}J_{i,j,r}(l) & \\
0, & \text{otherwise} & \\
\end{matrix}
\end{align}
where $J_{i,j,r}(l)=Wa_{i,j,r}(l)(1+\beta _{i,j}(l))\times$\\
$\biggl \lbrace \log _{2}(\frac{P_{i,j,r}^{ }(l)g_{i,j,r}(l)}{\tilde{I}_{i,j,r}+\frac{N_{p}}{\varpi 
_{i,j,r}^{I}(l)}+WN_{0}})-\frac{(\tilde{I}_{i,j,r}+WN_{0})\varpi _{i,j,r}^{I}(l)}{\ln (2)((\tilde{I}_{i,j,r}+WN_{0})\varpi _{i,j,r}^{I}(l)+N_{P})}\biggr \rbrace$
$+Wb_{i,j,r}(l)(1+\beta _{i,j}(l))+\delta 
_{i,j,r}(l)-\tau _{i,r}(l)$.

Subsequently, we have used a sub-gradient method to iteratively update the Lagrangian variables \cite{ali2019efficient} as
\begin{align}
\beta _{i,j}(l+1)&=\biggl[{\beta _{i,j}(l)+\zeta(l) \times \biggl \lbrace \tilde{\Pi }_{i,j}(l)-\Pi _{i,j}^{\min }(l)\biggr \rbrace }\biggr]^+ \\
\delta _{i,j,r}(l+1)&=\biggl[\delta _{i,j,r}(l)+\zeta(l) \times \lbrace {\sigma}_{i,j,r}(l)-\hat{\varpi }_{i,j,r}^{I}(l)\nonumber \\
&-\hat{\varpi }_{i,j,r}^{E}(l)\rbrace \biggr]^+,\\
{\tau _{i,r}(l+1)}&=\biggl[{\tau _{i,r}(l)}+\zeta(l) \times \biggl \lbrace 1-\sum_{j=1}^{J_{i}(l)}{{\sigma}_{i,j,r}(l)}\biggr \rbrace \biggr]^+,\\
{\theta _{i,j}(l+1)}&=\biggl[{\theta _{i,j}(l)}+\zeta(l) \times \biggl \lbrace \frac{\tilde{C}_{i,j}(l)}{\eta \hat{\varpi }_{i,j,r}^{E}}-\frac{P_{i,j}^{\min }(l)}{\eta \hat{\varpi }_{i,j,r}^{E}(l)}\biggr \rbrace \biggr]^+,\\
{\lambda _{i}(l+1)}&=\biggl[{\lambda _{i}(l)}+\zeta(l) \times \biggl \lbrace P_{i}^{T}-\sum_{j=1}^{J_{i}(l)}{\sum_{r=1}^{B}{\tilde{P}_{i,j,r}^{ }(l)}}\biggr \rbrace \biggr]^+,\\
{\pi _{i,j}(l+1)}&=\biggl[{\pi _{i,j}(l)}+\zeta(l) \times  \biggl \lbrace \frac{P^{RO}}{\eta \hat{\varpi }_{i,j,r}^{E}}-\frac{\tilde{C}_{i,j}(l)}{\eta \hat{\varpi }_{i,j,r}^{E}(l)}\biggr \rbrace \biggr]^+,
\end{align}
where $\zeta \geq 0$ denotes the step size and $[.]^+=\max\{0,.\}$. During each iteration, the value of variables are updated and the iterative process continues until convergence.

\section{Performance Evaluation}
\begin{figure*}[!htp]
\centering
\begin{tabular}{c}
\includegraphics[trim={0 0cm 0 0cm},clip,scale=.22]{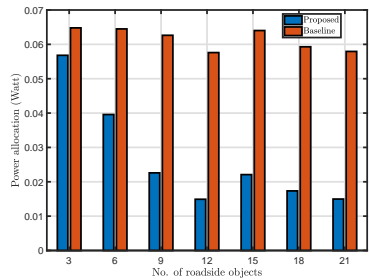} \\
(a) \\
\includegraphics[trim={0 0cm 0 0cm},clip,scale=.22]{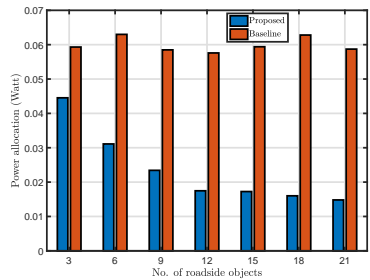} \\
(b) \\
\includegraphics[trim={0 0cm 0 0cm},clip,scale=.22]{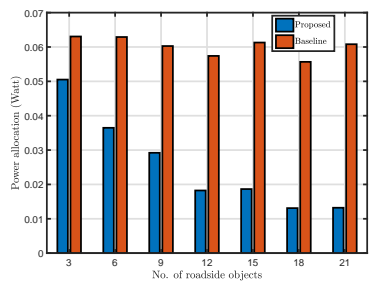} \\
(c)
\end{tabular}
\caption{Power allocation for the proposed and baseline techniques, where (a) vehicle velocity = 5 m/s, (b) vehicle velocity = 10 m/s, (c) vehicle velocity = 15 m/s.}
\label{fig_2_5}
\end{figure*}

Fig \ref{fig_2_5} compares the power allocation performance of the proposed optimization against the baseline technique. Generally, it can be seen that the proposed technique efficiently allocates power among roadside objects. \textcolor{black}{Particularly, we note that the power allocation level changes with an increase in the number of roadside objects. This is because more power is distributed when there are multiple roadside objects in the networks.} Moreover, to illustrate the impact of the velocity of electric vehicles against on power allocation, we plot Fig \ref{fig_2_5} (a), (b), and (c) when the velocity is 5 m/s, 10 m/s and 15 m/s, respectively. As shown in the figures, the power allocation levels drop with an increase in the speed of the vehicle. \textcolor{black}{Yet, we observe that the proposed optimization framework continues to allocate power efficiently and dynamically as opposed to the passive approach of the baseline technique.}

\begin{figure}[!htp]
\centering
\begin{tabular}{c}
\includegraphics[trim={0 0cm 0 0cm},clip,scale=.32]{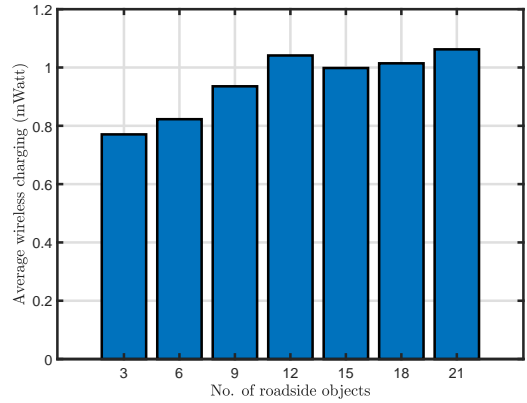}
\end{tabular}
\caption{Average wireless charging against different number of roadside objects.}
\label{fig_2_1}
\end{figure}

Fig \ref{fig_2_1} shows the average amount of wireless charging for the different number of roadside objects in the V2X network. In this case, the number of roadside objects connected to each vehicle is considered identical for each experiment. The number of roadside objects is incrementally increased from 3 to 21. In general, it can be noted that the average wireless charging increases with an increase in the number of roadside objects. However, as the number of connected roadside objects increases significantly, the amount of power consumed by each roadside object also increases. Thus, we observe a negligible improvement in wireless charging for a larger number of roadside objects.
 
\begin{figure}[!htp]
\centering
\begin{tabular}{c}
\includegraphics[trim={0 0cm 0 0cm},clip,scale=.32]{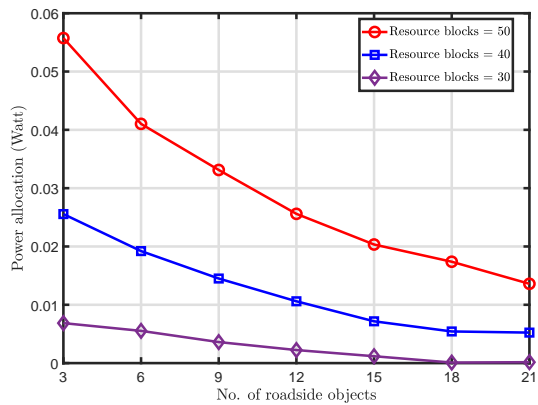}
\end{tabular}
\caption{Power allocation as a function of number of roadside objects for different resource blocks.}
\label{fig_2_2}
\end{figure}

Fig \ref{fig_2_2} shows the amount of allocated power against a different number of roadside objects. It can be noted that an increase in the number of roadside objects results in a reduction in the amount of allocated power. This dynamicity is one of the key features of the proposed optimization framework which takes into account different quality of service constraints allocating transmission power. Moreover, it can be observed that with an increase in the available resource blocks, the amount of allocated power also increases. By contrast, when the resource blocks are scarce, the number of allocated power drops. This trend can be attributed to the fact the distribution of power becomes much easier when there are a significantly larger number of available resource blocks. However, as evident from the figure, the impact of resource blocks diminishes for a larger number of roadside objects.  

\begin{figure}[!htp]
\centering
\begin{tabular}{c}
\includegraphics[trim={0 0cm 0 0cm},clip,scale=.32]{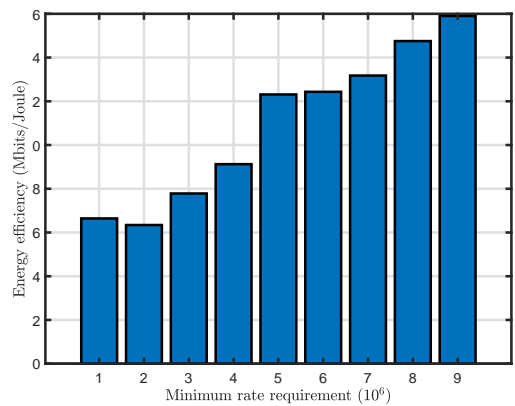}
\end{tabular}
\caption{Energy efficiency versus different minimum rate requirements.}
\label{fig_2_3}
\end{figure}

In Fig. \ref{fig_2_3}, we illustrate the achievable energy efficiency levels for different traffic loads. It can be seen that energy efficiency generally increases with an increase in the rate requirements from the roadside objects. In this regard, it is worth mentioning that the total number of connected roadside objects with each electric vehicle was kept at three. \textcolor{black}{As evident from the figure, our proposed optimization framework can improve energy efficiency under more stringent minimum rate requirements.}

\begin{figure}[!htp]
\centering
\begin{tabular}{c}
\includegraphics[trim={0 0cm 0 0cm},clip,scale=.32]{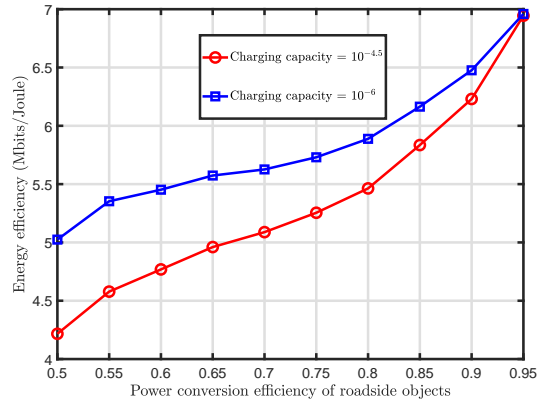}
\end{tabular}
\caption{Energy efficiency versus different charging capacity of the roadside object.}
\label{fig_2_4}
\end{figure}

Fig \ref{fig_2_4} shows energy efficiency against the power conversion efficiency of roadside objects. It can be seen from the figure that the total energy efficiency increases with an increase in power conversion efficiency. Moreover, the charging capacity, which indicates the required level of wireless charging for the operation of roadside objects, appears to have a profound effect on the overall energy efficiency. Specifically, when the charging capacity is low, the energy efficiency increases. By contrast, for a large value of the charging capacity, the energy efficiency is very low. This impact is much more vivid for smaller values of the power conversion efficiency, i.e, 0.5 to 0.8. However, the impact of charging capacity diminishes when the power conversion efficiency increases significantly.

\textcolor{black}{The results obtained in this section are critically important from a practical perspective. The C-V2X communications often require efficient allocation of scarce resources among the vehicles and roadside objects. The results provided here focus not only on the allocation of power but also on resource blocks. The results show that the proposed approach distributes power efficiently, thereby, reducing the energy budget of communication without degradation of the quality of service. This would also allow roadside objects to operate on the wirelessly transmitted power, thus, avoiding the need to charge manually through batteries. Finally,  the proposed optimization framework can improve energy efficiency under higher minimum rate requirements which also demonstrates the feasibility of the proposed approach for vehicular networks.}
 
\section{Conclusion and Future Work}

With the rapid evolution of cellular V2X networks, it is becoming ever more important to optimize the energy budget of such networks. With this motivation, this paper has developed a dynamic wireless-power transmission and resource allocation technique for cellular V2X networks. The formulated energy efficiency problem not only finds the best tradeoff between the power and capacity but also takes into account the quality of service requirements of electric vehicles and roadside objects. The proposed optimization framework optimizes the PS factors of roadside objects along with the power allocation resource block assignment from the electric vehicle. The results demonstrate the feasibility of the proposed solution to dynamically adjust the power levels and resources against the baseline technique.

Though the proposed solution provides considerable performance gains, it can be improved in a number of ways. For instance, the proposed solution can consider cooperation among vehicles to more efficiently allocate the resources and adjust PS factors. Besides this, multiple antenna links can be considered at the roadside objects or electric vehicles. In this way, the throughput can be further improved with the help of proper beamforming techniques. These challenging yet interesting extensions are left for future work. 

\section*{Acknowledgment}

This work was supported by the Finnish public funding agency for research, Business Finland under the project 5G Finnish Open Research Collaboration Ecosystem (5G-FORCE) which is part of 5G Test Network Finland (5GTNF).

\ifCLASSOPTIONcaptionsoff
  \newpage
\fi

\bibliographystyle{IEEEtran}
\bibliography{Wali_EE}
\begin{IEEEbiography}[{\includegraphics[width=1in,height=1.25in,clip,keepaspectratio]{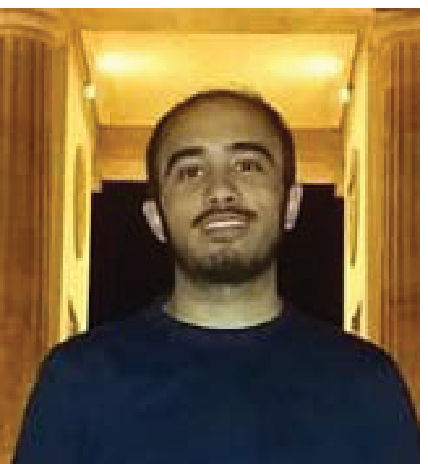}}]{Furqan Jameel} received his B.S. in Electrical Engineering (under ICT R\&D funded Program) in 2013 from the Lahore Campus of COMSATS Institute of Information Technology (CIIT), Pakistan. In 2017, he received his master's degree in Electrical Engineering (funded by prestigious Higher Education Commission Scholarship) at the Islamabad Campus of CIIT. In September 2018, he visited Simula Research Laboratory and the University of Oslo, Norway. From 2018 to 2019, he was with the University of Jyv\"askyl\"a, Finland, and Nokia Bell Labs, Espoo, where he worked as a researcher and a summer trainee, respectively. Currently, he is with the Department of Communications and Networking, Aalto University, Finland, where his research interests include modeling and performance enhancement of vehicular networks, machine/ deep learning, ambient backscatter communications, and wireless power transfer. He is the recipient of outstanding reviewer award 2017 from Elsevier.
\end{IEEEbiography}

\begin{IEEEbiography}[{\includegraphics[width=1in,height=1.25in,clip,keepaspectratio]{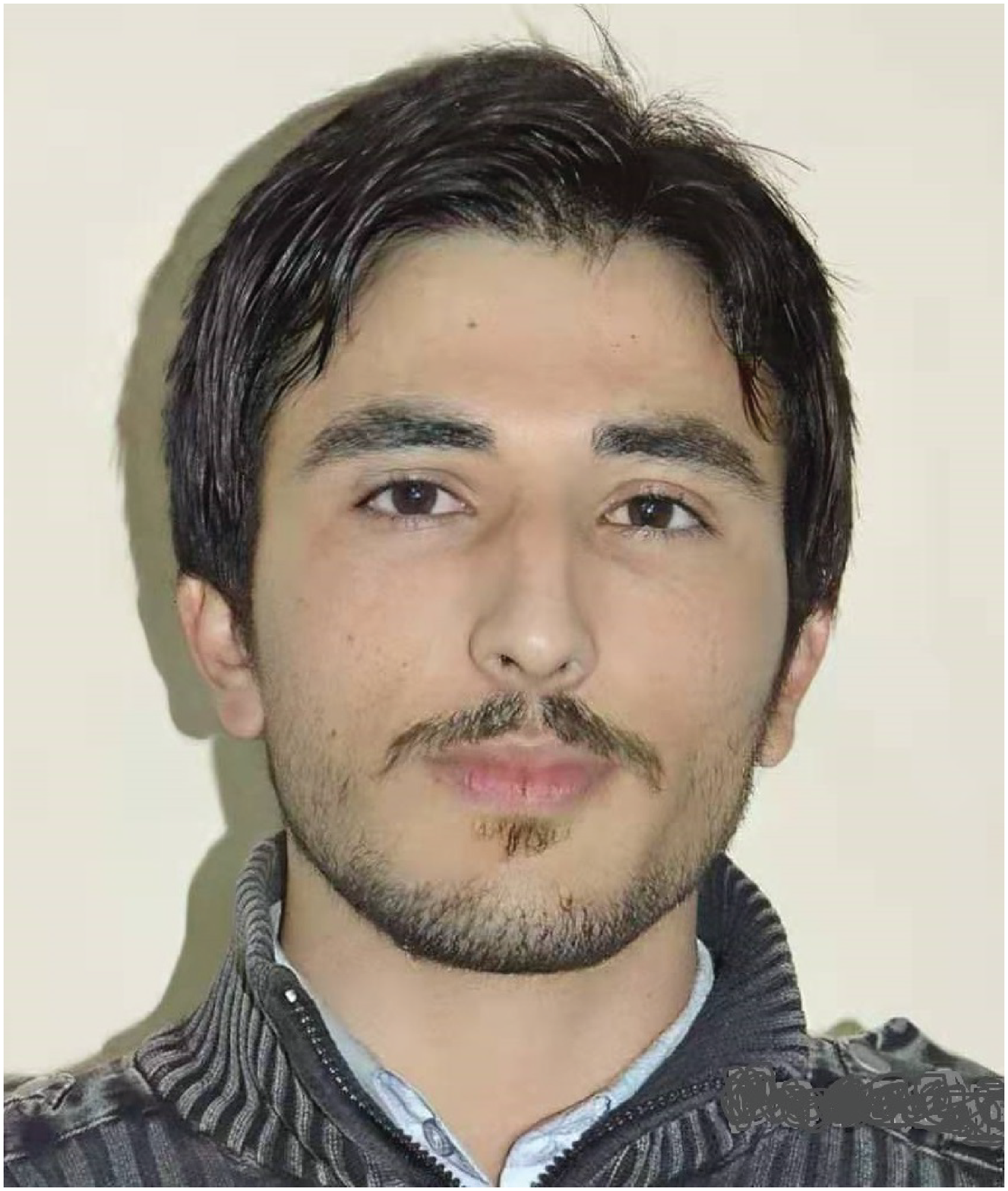}}]{Wali Ullah Khan} (Student Member, IEEE) received the B.S. degree (Hons.) in telecommunication from the University of Science and Technology Bannu, Khyber Pakhtunkhwa, Pakistan, in 2014, and the master's degree in electrical engineering from the Islamabad Campus of COMSATS Institute of Information Technology (currently known COMSATS University), Pakistan, in 2017. He is currently with the School of Information Science and Engineering, Shandong University, Qingdao, China. His research interests include convex/ non-convex optimization, non-orthogonal multiple access, energy/ spectral efficiency, physical layer security, heterogeneous networks, vehicular networks, machine/deep learning, and backscatter communications. He is also an Active Reviewer of several SCI journals, such as the IEEE WIRELESS COMMUNICATIONS, EURASIP Journal on Wireless Communications and Networking, IEEE ACCESS, the IEEE OPEN JOURNAL OF VEHICULAR TECHNOLOGY, the IEEE TRANSACTIONS ON INDUSTRIAL INFORMATICS, the Internet Technology Letters, and Physical Communication
\end{IEEEbiography}

\begin{IEEEbiography}[{\includegraphics[width=1in,height=1.25in,clip,keepaspectratio]{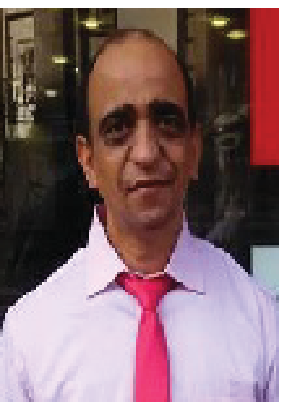}}]{Neeraj Kumar} (M'17, SM'17) received his Ph.D. in CSE from Shri Mata Vaishno Devi University, Katra (J \& K), India in 2009, and was a postdoctoral
research fellow in Coventry University, Coventry, UK. He is working as an Associate Professor in the Department of Computer Science
and Engineering, Thapar Institute of Engineering and Technology (Deemed to be University), Patiala (Pb.), India. He has published more than 200 technical research papers in top cited journals such as IEEE TKDE, IEEE TIE, IEEE TDSC, IEEE TITS, IEEE TCE, IEEE TII, IEEE TVT, IEEE ITS, IEEE SG, IEEE Netw., IEEE Comm., IEEE WC, IEEE IoTJ, IEEE SJ, Computer Networks, Information sciences, FGCS, JNCA, JPDC and ComCom. He has guided many research scholars leading to Ph.D. and M.E./M.Tech. His research is supported by funding from UGC, DST, CSIR, and TCS. He is an Associate Technical Editor of IEEE Communication Magazine. He is an Associate Editor of IJCS, Wiley, JNCA, Elsevier, and Security \& Communication, Wiley. He is senior member of the IEEE.
\end{IEEEbiography}

\begin{IEEEbiography}[{\includegraphics[width=1in,height=1.25in,clip,keepaspectratio]{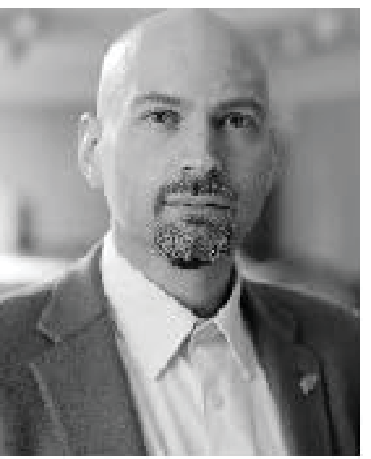}}]{Riku J\"antti} received the M.Sc. degree (Hons.) in electrical engineering and the D.Sc. degree (Hons.) in automation and systems technology from the Helsinki University of Technology (TKK), in 1997 and 2001, respectively. He was a Professor pro term with the Department of Computer Science, University of Vaasa. In 2006, he joined the School of Electrical Engineering, Aalto University (formerly known as TKK), Finland, where he is currently a Professor in communications engineering and the Head of the Department of Communications and Networking. His research interests include radio resource control and optimization for machine type communications, cloud-based radio access networks, spectrum and co-existence management, and RF inference. He is an Associate Editor of the IEEE TRANSACTIONS ON VEHICULAR TECHNOLOGY. He is also an IEEE VTS Distinguished Lecturer (Class 2016).
\end{IEEEbiography}

\end{document}